\documentclass[twocolumn,english,prl,notitlepage,nofootinbib,floatfix,superscriptaddress]{revtex4-2}
\usepackage[T1]{fontenc}
\usepackage[utf8]{inputenc}
\usepackage{refstyle}
\usepackage{amsmath}
\usepackage{bm} 
\usepackage{amssymb}
\usepackage{graphicx}
\usepackage[pdftex, hidelinks]{hyperref}
\usepackage[usenames,dvipsnames]{xcolor}
\usepackage{bbm}
\usepackage{float}
\usepackage{tikz}
\usepackage{xspace}
\usepackage{bm}
\usepackage{booktabs}
\usepackage{babel}
\usepackage[separate-uncertainty=true]{siunitx}
\usepackage[makeroom]{cancel}
\usepackage{siunitx}
\usepackage{amssymb}
\usepackage[normalem]{ulem} 
\usepackage{comment}  
\usepackage{braket}  
\usepackage{float} 
\usepackage{amsmath}
\usetikzlibrary{matrix}

\newcommand\myshade{85}
\colorlet{mylinkcolor}{violet}
\colorlet{mycitecolor}{YellowOrange}
\colorlet{myurlcolor}{Aquamarine}
\hypersetup{
  linkcolor  = mylinkcolor!\myshade!black,
  citecolor  = mycitecolor!\myshade!black,
  urlcolor   = myurlcolor!\myshade!black,
  colorlinks = true,
}

\usepackage{soul}
\usepackage{xr}

\DeclareMathAlphabet\mathbfcal{OMS}{cmsy}{b}{n}

\begin{document}

\title{Preselection-Encoded Weak-Value Interferometry for Optical-Rotation Metrology}

\author{Jinyi Li}
\thanks{These authors contributed equally to this work.}
\affiliation{School of Physics and Optical Engineering, Zhejiang University of Technology, Hangzhou 310023, China}
\affiliation{Laboratory of Spin Magnetic Resonance, School of Physical Sciences, Anhui Province Key Laboratory of Scientific Instrument Development and Application, University of Science and Technology of China, Hefei 230026, China}

\author{Juncheng Zheng}
\thanks{These authors contributed equally to this work.}
\affiliation{School of Physics and Optical Engineering, Zhejiang University of Technology, Hangzhou 310023, China}

\author{Yuntao Shen}
\affiliation{School of Physics and Optical Engineering, Zhejiang University of Technology, Hangzhou 310023, China}

\author{Jiashu Liang}
\affiliation{School of Physics and Optical Engineering, Zhejiang University of Technology, Hangzhou 310023, China}

\author{Sicheng Pan}
\affiliation{School of Physics and Optical Engineering, Zhejiang University of Technology, Hangzhou 310023, China}

\author{Dongmei Li}
\affiliation{School of Physics and Optical Engineering, Zhejiang University of Technology, Hangzhou 310023, China}

\author{Hengyan Wang}
\email{hywang@zust.edu.cn}
\affiliation{Department of Physics, Zhejiang University of Science and Technology, Hangzhou 310023, China}

\author{Wenqiang Zheng}
\email{wqzheng@zjut.edu.cn}
\affiliation{School of Physics and Optical Engineering, Zhejiang University of Technology, Hangzhou 310023, China}

\begin{abstract} 
Optical-rotation measurements benefit from increasing probe photon number, but their performance is often limited by detector saturation and technical noise in the optical readout. Here we introduce preselection-encoded weak-value amplification (PeWVA), in which the quantity of interest is encoded into the probe polarization before a downstream weak-value interferometer. This architecture separates the sensing interaction from the postselected readout, enabling tunable local amplification while reducing the optical power delivered to the detector. We show that PeWVA does not create additional Fisher information, but offers a practical advantage when detector dynamic range is limiting. We further analyze coherent stray light and phase fluctuations and show reduced conversion of these perturbations into equivalent rotation errors. Using an $^{87}$Rb $M_x$ atomic magnetometer, we experimentally demonstrate substantially improved magnetic-field readout under detector-constrained conditions using both transimpedance-amplified and avalanche photodiodes. The experiments also confirm reduced susceptibility to vapor-cell-induced stray light and interferometer phase modulation. These results establish PeWVA as a detector-constrained optical readout strategy with enhanced tolerance to representative interferometric perturbations.

\end{abstract}

\maketitle

\section{introduction}  

Optical rotation (OR), the rotation of the polarization plane of linearly polarized light induced by interaction with a physical medium, is a widely used transduction mechanism in precision measurement. It provides an optical interface to physical quantities that are otherwise difficult to access directly. Atomic magnetometry is one representative example, in which a magnetic field modifies the collective atomic spin and the spin response is mapped onto the polarization of a probe through Faraday rotation \cite{budker2007optical,dang2010ultrahigh}. Closely related OR readouts are used in atomic gyroscopes \cite{kornack2005nuclear},  magnetic-domain measurements \cite{mccord2015progress}, chiral molecular measurements \cite{delage2025chiral}, and astrophysical Faraday-rotation studies \cite{gaensler2005magnetic}. Across these systems, the central readout problem is the precise estimation of a small polarization change.

For coherent optical probing, increasing the photon number $N_{\mathrm{ph}}$ is a direct route to improving measurement precision. In the ideal photon-shot-noise limit, the parameter uncertainty scales as $1/\sqrt{N_{\mathrm{ph}}}$, while the corresponding Fisher information scales linearly with $N_{\mathrm{ph}}$ \cite{kay1993fundamentals,braunstein1994statistical,demkowicz2015quantum}. The same principle motivates the use of high optical power in many precision measurements. Gravitational-wave interferometers, for example, employ very large circulating optical powers \cite{smith2009path}, while it is also an important resource in gyroscopy, optical-frequency metrology, and optical coherence tomography \cite{hokmabadi2019non,udem2002optical,huang1991optical}. In practice, however, the optical power incident on the detector cannot be increased without bound because photodetectors have a finite linear dynamic range and eventually exhibit saturation and nonlinear response. This limitation can be particularly severe for low-noise or high-gain detectors, whose favorable noise performance is often available only over a relatively low optical-power range. The resulting challenge is to use a large photon flux to interrogate the sensing system while keeping the detected optical power within the linear operating range of the photodetector. This distinction between the photon number available at the sensing stage and that received by the detector is especially relevant for avalanche photodiodes (APDs), single-photon detectors, and other high-gain receivers that can offer excellent noise performance but saturate at comparatively low optical power. It therefore motivates a readout strategy that can preserve the benefit of a large incident photon number while substantially reducing the optical power delivered to the detector.

Weak-value amplification (WVA) exploits nearly orthogonal pre- and postselection to convert a small parameter-dependent perturbation into an enhanced change of the detected signal, and has been demonstrated in a wide range of precision-measurement settings \cite{aharonov1988result,jordan2014technical,feizpour2011amplifying,viza2015experimentally,xu2013phase,hosten2008observation,dixon2009ultrasensitive,kocsis2011observing,viza2013weak,magana2014amplification}. A characteristic feature of WVA is that this enhanced response can be obtained while only a small fraction of the incident optical power reaches the detector, making it attractive for detector-constrained metrology.
Under ideal detection and fixed incident resources, postselection does not create additional Fisher information, but rather redistributes the available information among the measurement outcomes \cite{zhang2015precision,tanaka2013information,knee2014amplification}. Its practical advantage instead appears when the measurement is limited by detector nonidealities or specific technical-noise mechanisms, including saturation, finite dynamic range, digitization, and detector noise \cite{jordan2014technical,harris2017weak,xu2020approaching}. The practical role of WVA in the present work is to relax the detector constraint on the usable optical power. A larger incident photon number can improve the upstream optical statistics and signal-to-noise ratio (SNR), while postselection reduces the photon flux delivered to the detector. In this way, the sensing stage can benefit from a stronger optical probe without necessarily driving the detector into saturation. This separation between the photon resources available at the sensing stage and the photon flux tolerated by the detector is central to the measurement strategy developed below.

An additional consideration is the readout modality of the weak-value signal. In many WVA implementations, a small parameter is mapped onto a transverse displacement or centroid shift of an optical beam and subsequently retrieved using spatially resolved detection, such as a CCD \cite{dixon2009ultrasensitive,qiu2016estimation}. Related weak-measurement readouts may also rely on spectrally resolved detection \cite{li2016application}. The acquisition rate of such camera- or spectrometer-based readout can limit the measurement of rapidly varying signals. Polarization interferometry provides an alternative approach by converting a small polarization rotation directly into an output-intensity variation, which can be detected with a fast photodetector. Interferometric weak measurements of photon polarization have also demonstrated the connection between path interference, measurement resolution, and measurement back-action \cite{iinuma2011weak}. We therefore combine polarization interferometry with postselected weak-value readout, enabling OR signals to be retrieved directly from photodiode intensity measurements.

A further difficulty arises when the parameter-dependent sensing interaction itself is embedded in the weak-measurement interferometer. For a general sensing medium, absorption, scattering, refractive-index fluctuations, and thermal or mechanical perturbations can then be converted directly into interferometric noise. The issue is especially transparent in an atomic vapor cell, where absorption changes the arm powers, multiple reflections from the glass windows generate coherent stray light, and cell heating or air convection can perturb the optical phase and spatial mode. Stray light is a recognized limitation in precision interferometry \cite{sasso2019lisa}, and multiple reflections from an atomic cell can also limit weak-value atomic-magnetometer readout \cite{lin2023high}. Here we introduce preselection-encoded weak-value amplification (PeWVA), in which the physical sensing interaction and the postselected interferometric readout are separated. The sensing medium first converts the quantity of interest into a polarization rotation of the probe. This parameter-dependent polarization state becomes the preselected state of a downstream interferometer. The weak system-pointer coupling inside the interferometer is fixed by optical elements and is therefore independent of the unknown parameter. In contrast to conventional parameter-encoded WVA, the information is encoded in the preselection rather than in the weak interaction itself.

We use an $^{87}\mathrm{Rb}$ M${_\mathrm{x}}$ atomic magnetometer as a controlled platform to demonstrate the PeWVA readout architecture. The magnetometer maps the magnetic-field-dependent atomic-spin response onto a time-dependent Faraday rotation, while the heated and optically imperfect vapor cell provides a stringent test of the benefit of separating the sensing interaction from the downstream interferometric readout. The key question is whether the metrological advantages associated with conventional WVA, particularly its response to detector constraints and technical perturbations, survive when the parameter is encoded in the preselected state rather than in the weak interaction. We address this question by deriving the PeWVA response and Fisher information, analyzing detector saturation, coherent stray light, and interferometer phase jitter, and experimentally comparing PeWVA with standard interferometric (SI) readout using both a transimpedance-amplified photodiode (PD) and an avalanche photodiode (APD). 


\section{THEORETICAL FRAMEWORK}

PeWVA differs from conventional parameter-encoded weak-value amplification in where the unknown parameter enters the measurement chain. In conventional WVA, the quantity of interest is usually encoded in the weak system--pointer interaction. In PeWVA, by contrast, the sensing process first prepares a parameter-dependent optical state, whereas the downstream weak coupling is fixed and independently controlled. We therefore formulate the theory first for a generic polarization-encoded signal, without specifying the physical origin of the encoding. This separation raises a central question of the present work: whether the metrological advantages established for conventional WVA remain valid when the parameter encoding is transferred from the weak interaction to the preselected state. The atomic Faraday-rotation implementation is then introduced as a specific realization of the general framework.

\subsection{General preselection-encoded weak-value framework}
 
Let the sensing stage prepare a normalized polarization state
\begin{equation}
\ket{\chi(\theta)}
=
\sin\theta\,\ket{H}
+
\cos\theta\,\ket{V},
\label{eq:generic_state}
\end{equation}
where $\ket{H}$ and $\ket{V}$ denote the horizontal and vertical linear polarization states, respectively, and $\theta$ is the parameter-dependent polarization angle. More generally,
\begin{equation}
\theta=\theta_0+\Delta\theta(q),
\label{eq:generic_theta}
\end{equation}
where $q$ denotes the physical quantity of interest and $\theta_0$ is an experimentally adjustable bias. The defining feature of PeWVA is that $q$ is already encoded in the preselected state before the weak interaction.

\begin{figure}[t]
\centering
\includegraphics[width=0.92\linewidth]{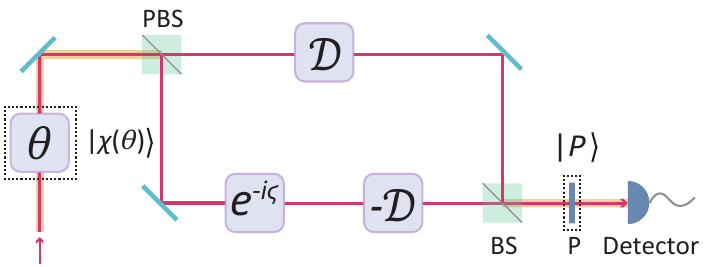}
\caption{Schematic of the PeWVA polarization-interferometric readout. The sensing stage prepares the parameter-dependent state $\ket{\chi(\theta)}$ before the interferometer. The opposite polarization rotations $\pm\mathcal{D}$ provide the fixed weak coupling, $\varsigma$ denotes the relative phase between the two interferometer arms, and the final polarizer performs the postselection onto $\ket{P}$. Removing the final polarizer gives the standard-interferometric (SI) reference. PBS, polarizing beam splitter; BS, beam splitter.}
\label{weak measurement schematic diagram}
\end{figure}

The state $\ket{\chi(\theta)}$ is then sent to the polarization interferometer shown in Fig.~\ref{weak measurement schematic diagram}. A polarizing beam splitter (PBS) separates the horizontal and vertical polarization components into two paths. Opposite small polarization rotations $\pm\mathcal{D}$ are introduced in the two arms and constitute a fixed weak system--pointer coupling. A controllable relative phase $\varsigma=2n\pi$, where $n$ is an integer, is introduced in one arm, after which the two paths are recombined at a 50/50 beam splitter (BS). Neglecting phase perturbations for the moment, the transformations associated with the two output ports can be written as
\begin{equation}
U^{\pm}
=
\frac{1}{\sqrt{2}}
\left[
 e^{-i2\mathcal{D}\hat A}\ket{H}\bra{H}
 \pm
 e^{i2\mathcal{D}\hat A}\ket{V}\bra{V}
\right],
\label{eq:U_pm}
\end{equation}
where $\hat{A}=\left| R \right> \left< R \right|-\left| L \right> \left< L \right|$,
with $\ket{R}$ and $\ket{L}$ denoting right- and left-circular polarization states, respectively. Thus, $\mathcal{D}$ is a fixed and controllable coupling parameter, whereas the measured quantity appears through $\theta$.

A polarizer placed before the detector postselects the field onto $\left| P \right> =\left( \left| L \right> +\left| R \right> \right) /\sqrt{2}$. The corresponding weak value is
\begin{equation}
A_w
=
\frac{\bra{P}\hat A\ket{\chi}}
{\langle P|\chi\rangle}
=
-i\cot\theta.
\label{eq:weak_value_general}
\end{equation}
Hence $|A_w|$ becomes large as the preselected and postselected states approach orthogonality.
In PeWVA, the amplified response arises from the parameter dependence of the preselected state, whereas the weak coupling $\mathcal{D}$ remains fixed and independently controlled.

After the interferometer, the states at the constructive $(+)$ and destructive $(-)$ output ports are
\begin{align}
\ket{\chi_{out}^{\pm}}
&=U^{\pm}\ket{\chi}
\nonumber\\
&=
\frac{1}{\sqrt{2}}
\left(
\sin\theta\cos2\mathcal{D}
\pm
\cos\theta\sin2\mathcal{D}
\right)\ket{H}
\nonumber\\
&\quad+
\frac{1}{\sqrt{2}}
\left(
\sin\theta\sin2\mathcal{D}
\pm
\cos\theta\cos2\mathcal{D}
\right)\ket{V}.
\label{eq:output_state_general}
\end{align}
After postselection, the detected intensities are
\begin{align}
I_P^{\pm}(\theta)
&=
I_0\left|\bra{P}U^{\pm}\ket{\chi}\right|^2
\nonumber\\
&=
\frac{I_0}{2}
\sin^2\!\left(\theta\pm2\mathcal{D}\right),
\label{eq:PeWVA_intensity_general}
\end{align}
where $I_0$ is the incident optical intensity. In the weak-value limit, $|A_w|\gg 1$, Eq.~(\ref{eq:PeWVA_intensity_general}) can be recast as
\begin{equation}
I_P^{\pm}(\theta)
\approx
\frac{I_0}{2|A_w|^2}
\left[
\cos2\mathcal{D}
\mp
\operatorname{Im}(A_w)\sin2\mathcal{D}
\right]^2.
\label{eq:PeWVA_intensity_weakvalue}
\end{equation}

The same interferometer provides an SI reference simply by removing the final postselection polarizer. The two output intensities are then
\begin{equation}
I_{\mathrm{SI}}^{\pm}(\theta)
=
\frac{I_0}{2}
\left[
1\pm\sin(2\theta)\sin(4\mathcal{D})
\right].
\label{eq:SI_intensity_general}
\end{equation}
Thus, PeWVA and SI read out the same upstream polarization signal through different transfer functions.
 
To compare their local responses, we define the normalized visibility
\begin{equation}
\mathcal{V}
=
\frac{I^+-I^-}{I^++I^-}.
\label{eq:visibility_definition}
\end{equation}
For PeWVA and SI, respectively,
\begin{equation}
\mathcal{V}_{P}
=
\frac{\sin(2\theta)\sin(4\mathcal{D})}
{1-\cos(2\theta)\cos(4\mathcal{D})},
\label{eq:PeWVA_visibility}
\end{equation}
and
\begin{equation}
\mathcal{V}_{\mathrm{SI}}
=
\sin(2\theta)\sin(4\mathcal{D}),
\label{eq:SI_visibility}
\end{equation}
For SI, the response is maximized at $\mathcal{D}=22.5^\circ$, which is therefore used in the following comparison.

\begin{figure}[t]
\centering
\includegraphics[width=0.8\linewidth]{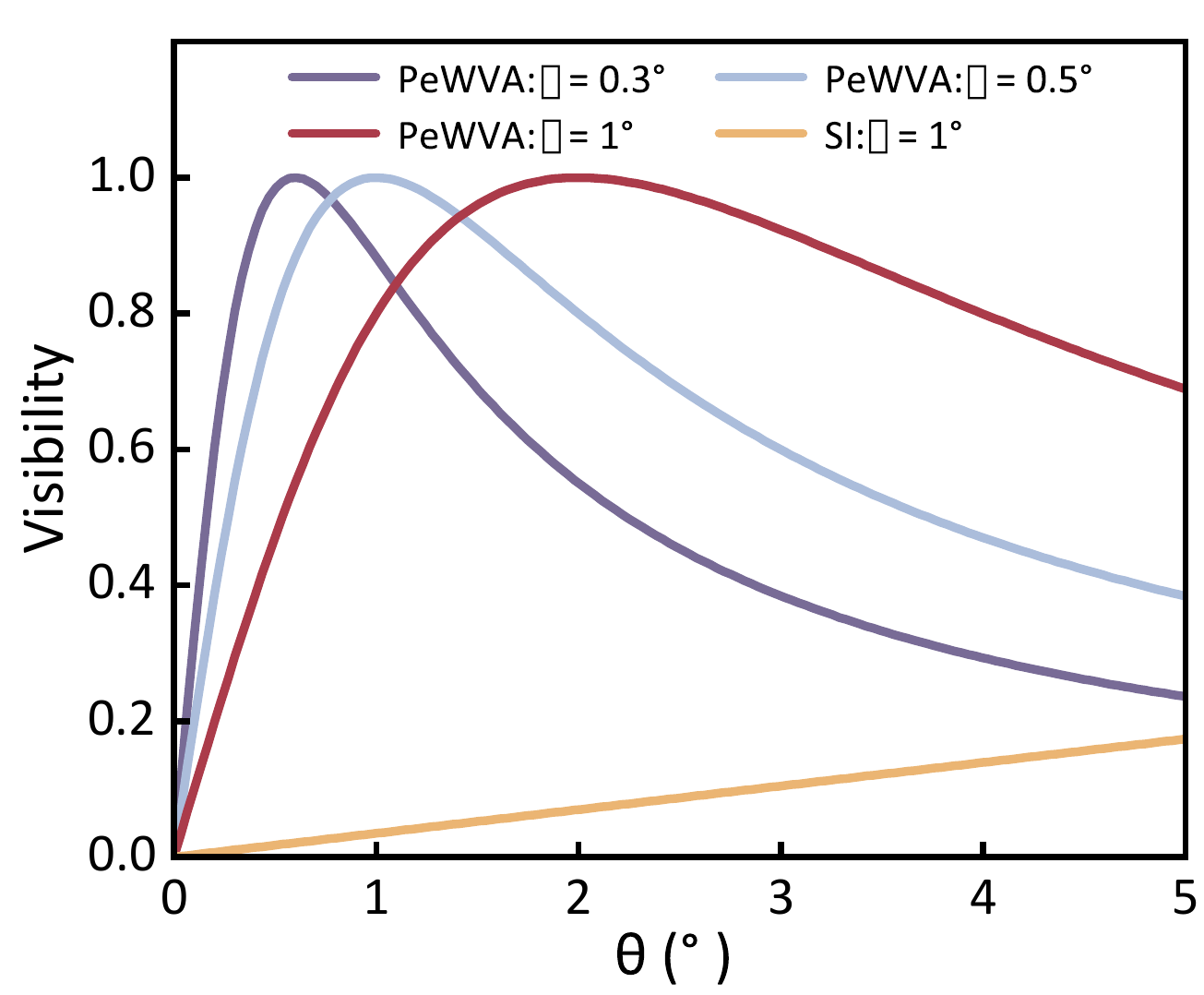}
\caption{Calculated visibility for SI and PeWVA as a function of the polarization angle $\theta$. For SI, $\mathcal{D}=22.5^\circ$ maximizes the local response. For PeWVA, decreasing $\mathcal{D}$ enhances the local response near the operating point while narrowing the usable range, illustrating the trade-off between local amplification and dynamic range.}
\label{simulation results}
\end{figure}

In the weak-coupling regime, $\mathcal{D}\ll1$, and close to the nearly orthogonal condition, $\theta\ll1$,
\begin{equation}
\mathcal{V}_{P}
\simeq
\frac{4\theta\mathcal{D}}
{\theta^2+4\mathcal{D}^2}.
\label{eq:PeWVA_visibility_small}
\end{equation}
Equation~(\ref{eq:PeWVA_visibility_small}) provides a direct picture of the 
gain--dynamic-range trade-off in PeWVA. In the vicinity of the nearly 
orthogonal pre- and postselection condition, where 
$\theta \ll 2\mathcal{D}$, the visibility reduces to
$\mathcal{V}_{P}\simeq \theta/\mathcal{D}$. The corresponding local slope is 
therefore approximately 
$\partial\mathcal{V}_{P}/\partial\theta\simeq1/\mathcal{D}$, showing explicitly 
that a smaller weak-coupling angle produces a larger response to a given small 
variation of the preselection angle. At the same time, decreasing $\mathcal{D}$ narrows the high-response region in $\theta$. In fact, Eq.~(\ref{eq:PeWVA_visibility_small}) reaches its 
maximum at $\theta\simeq2\mathcal{D}$, indicating that the region over which 
the response changes rapidly is shifted toward $\theta=0$ and becomes 
progressively narrower as $\mathcal{D}$ is reduced. As shown in Fig.~\ref{simulation results}, reducing $\mathcal{D}$ enhances the local response around the operating point at the expense of a narrower linear range in $\theta$. 
Therefore, the enhancement provided by a small $\mathcal{D}$ should be 
interpreted as a tunable local amplification rather than an unrestricted 
increase in sensitivity over the entire measurement range. The optimal value 
of $\mathcal{D}$ is consequently determined by the required compromise between 
local gain and usable dynamic range. Moreover, as the preselected and postselected states approach orthogonality, the response becomes increasingly nonlinear, further limiting the usable linear range and requiring higher-order weak-measurement effects to be taken into account \cite{nishizawa2015weak}.
 
\subsection{Fisher information under detector constraints}

The local slope enhancement discussed above does not by itself imply a larger amount of information per incident photon. To quantify the information available for estimating the polarization angle $\theta$, we use the classical Fisher information (FI). For $N$ independent incident photons with mutually exclusive outcome probabilities $p_i(\theta)$, the total FI is
\begin{equation}
F(\theta)
=
N\sum_i
\frac{1}{p_i(\theta)}
\left[
\frac{\partial p_i(\theta)}{\partial\theta}
\right]^2.
\label{eq:FI_definition}
\end{equation}
Physically, the FI quantifies how rapidly the measurement statistics change with the parameter to be estimated. A
larger FI therefore means that two nearby values of $\theta$ produce more distinguishable outcome distributions, and
hence permit a smaller statistical uncertainty in principle. For an unbiased estimator $\hat{\theta}$, the Cram\'er--Rao bound is $Var( \hat{\theta}) \ge 1/F\left( \theta \right)$ \cite{kay1993fundamentals,braunstein1994statistical}.

For SI, each incident photon exits through either the constructive or destructive output port with probabilities 
\begin{equation}
p_{\mathrm{SI}}^{\pm}
=
\frac{1}{2}
\left[
1\pm\sin(2\theta)\sin(4\mathcal{D})
\right],
\label{eq:SI_probabilities}
\end{equation}
Substituting these probabilities into Eq.~(\ref{eq:FI_definition}) gives
\begin{equation}
F_{\mathrm{SI}}
=
\frac{
4N\left[\cos(2\theta)\sin(4\mathcal{D})\right]^2
}{
1-\sin^2(2\theta)\sin^2(4\mathcal{D})
}.
\label{eq:FI_SI}
\end{equation}
The SI response is optimized at $\mathcal{D}=22.5^\circ$, for which the ideal FI reaches $4N$. At this setting, $\sin(4\mathcal{D})=1$, so the two-port SI
measurement extracts the maximum information available in this ideal polarization-interferometric model. This also provides the natural SI reference used in the following comparison.

For PeWVA, each incident photon has three possible outcomes: detection at the constructive port, detection at the destructive port, or rejection by postselection. The corresponding probabilities are
\begin{equation}
\begin{aligned}
p_+&=\frac{1}{2}\sin^2(\theta+2\mathcal{D}),\\
p_-&=\frac{1}{2}\sin^2(\theta-2\mathcal{D}),\\
p_0&=\frac{1}{2}
\left[
1+\cos(2\theta)\cos(4\mathcal{D})
\right].
\end{aligned}
\label{eq:PeWVA_probabilities}
\end{equation}
Using these three outcomes in Eq.~(\ref{eq:FI_definition}) yields
\begin{align}
F_P
=
2N\Bigg[
&
\cos^2(\theta+2\mathcal{D})
+
\cos^2(\theta-2\mathcal{D})
\nonumber\\
&+
\frac{
\sin^2(2\theta)\cos^2(4\mathcal{D})
}{
1+\cos(2\theta)\cos(4\mathcal{D})
}
\Bigg].
\label{eq:FI_PeWVA}
\end{align}
The three terms in Eq. (\ref{eq:FI_PeWVA}) account for the parameter dependence of the two detected outputs and of the postselection-rejection channel. The latter is essential in an information-theoretic comparison based on the complete set of
possible outcomes, because the rejection probability itself also varies with $\theta$.

\begin{figure}[t]
\centering
\includegraphics[width=0.92\linewidth]{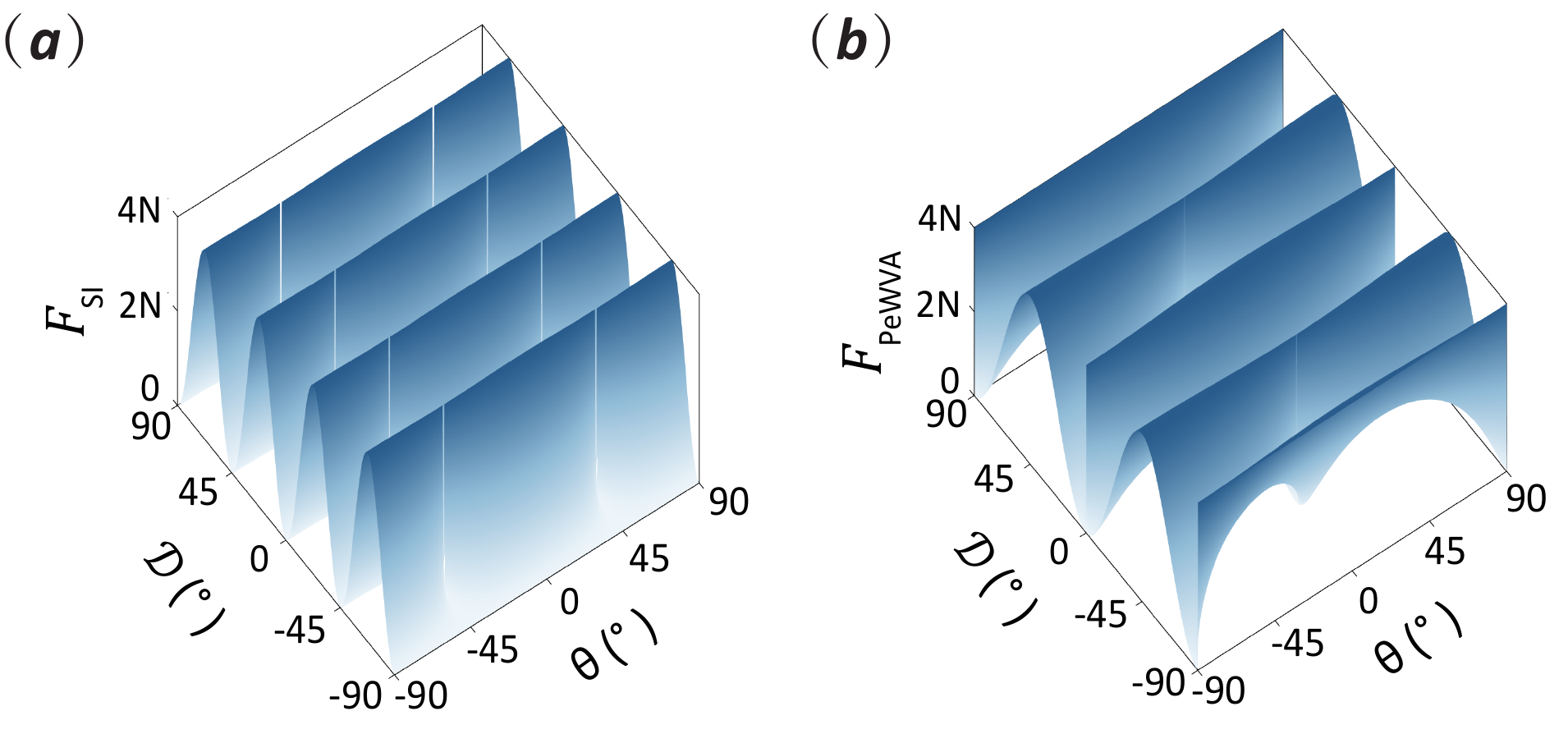}
\caption{Fisher information as a function of $\mathcal{D}$ and $\theta$ for (a) SI and (b) PeWVA. At fixed incident photon number, both schemes attain the same ideal maximum FI of $4N$.}
\label{fig:fisher information}
\end{figure}

As shown in Fig.~\ref{fig:fisher information}, both schemes can attain the same ideal maximum FI of $4N$ at fixed incident photon number, consistent with the general information-theoretic analysis of weak measurements \cite{zhang2015precision,demkowicz2015quantum,tanaka2013information,knee2014amplification}. The enhanced local response of PeWVA therefore represents a redistribution of the measurement response rather than an increase in the information carried by each incident photon. The FI landscapes also clarify the different roles of $\mathcal{D}$ in the two schemes. In SI, the optimum occurs at the strong-interference setting $\mathcal{D}=22.5^\circ$. In PeWVA, by contrast, the FI approaches $4N$ as $\mathcal{D}\rightarrow0$, even though the detected fraction becomes small. Thus, reducing $\mathcal{D}$ redistributes the same incident-photon information into a much lower-power postselected output.

The practical distinction emerges when the detector, rather than the incident optical field, is the limiting resource. The total PeWVA postselection probability is
\begin{equation}
p_s
=
p_++p_-
=
\frac{1}{2}
\left[
1-\cos(2\theta)\cos(4\mathcal{D})
\right].
\label{eq:postselection_probability}
\end{equation}
Near the weak-value operating condition, both $\theta$ and $\mathcal{D}$ are small and $p_s\ll1$. Consequently, only a small fraction of the photons that interrogate the sensing stage reach the detector, while the parameter-dependent preselected state is still prepared using the full incident probe field. 

If the detector remains linear only up to a total detected photon number $N_{\mathrm{sat}}$, SI can use an incident photon number of order $N_{\mathrm{sat}}$ under the same detector constraint, whereas PeWVA can accommodate $N_{\text{in}}\sim N_{\text{sat}}/p_s$ incident photons while keeping the postselected photon number near $N_{\mathrm{sat}}$. In the ideal weak-coupling limit $\mathcal{D}\rightarrow0$, $F_P=4N$ and $p_s = \sin^2\theta =
|\langle P|\chi\rangle|^2$.
The FI accessible before detector saturation therefore scales as
\begin{equation}
F_{P}^{(\mathrm{sat})}
\sim
\frac{4N_{\mathrm{sat}}}{p_s},
\label{eq:FI_saturation_scaling}
\end{equation}
compared with $F_{\mathrm{SI}}^{(\mathrm{sat})}\sim4N_{\mathrm{sat}}$ for SI under the same total detected-photon constraint. In this idealized limit the ratio of the detector-limited information resources scales as $F_P^{(\mathrm{sat})}/F_{\mathrm{SI}}^{(\mathrm{sat})}\sim1/p_s$. The enhancement can therefore become substantial when the postselection probability is small.
The scaling in Eq.~(\ref{eq:FI_saturation_scaling}) reflects an advantage that arises from the detector constraint, rather than from an increase in the information carried by each incident photon. By reducing the photon flux delivered to the detector, postselection allows a larger incident photon flux to interrogate the sensing stage before detector saturation occurs. This distinction is particularly relevant for high-gain or low-noise detectors with limited linear dynamic range. 

\subsection{Response to specific interferometric perturbations}

We next consider two representative perturbations of the interferometric readout, coherent stray light and relative phase fluctuations, and quantify the corresponding equivalent optical-rotation errors for PeWVA and SI at the operating points considered here.

\subsubsection{Coherent stray light}

Parasitic reflections from imperfect optical components can coherently overlap
with the signal field and introduce systematic errors in precision
interferometry \cite{sasso2019lisa}. Similar effects can also limit weak-value
measurements, particularly when multiple reflections are introduced by optical
windows or atomic vapor cells
\cite{dixon2009ultrasensitive,lin2023high}. We model the combined effect of these parasitic reflections as an effective coherent stray field
\begin{equation}
E_s = r e^{i\delta}E_{\mathrm{in}},
\label{eq:stray_field}
\end{equation}
where $E_{\mathrm{in}}$ denotes the incident optical field amplitude, $r$ is the relative amplitude of the stray field, and $\delta$ is its phase relative to the signal field.

For PeWVA, the normalized signal-field amplitude at the two output ports after postselection is
\begin{equation}
A_P^\pm(\theta)
=
\frac{1}{\sqrt{2}}
\sin(\theta\pm2\mathcal{D}).
\end{equation}
We assume that the representative stray component bypasses the weak
polarization rotation in the interferometer but is projected by the same
postselection polarizer. Its contribution at the detector is therefore
$r e^{i\delta}\sin\theta$. The total detected intensity becomes
\begin{align}
\widetilde I_P^\pm
&=
I_0
\left|
\frac{1}{\sqrt{2}}
\sin(\theta\pm2\mathcal{D})
+
r e^{i\delta}\sin\theta
\right|^2
\nonumber\\
&=
I_P^\pm
+
I_0
\left[
\sqrt{2}r\sin\theta
\sin(\theta\pm2\mathcal{D})\cos\delta
+
r^2\sin^2\theta
\right],
\label{eq:stray_PeWVA}
\end{align}
where $I_P^\pm = {I_0}\sin^2(\theta\pm2\mathcal{D})/2$ is the ideal PeWVA intensity. The stray-light-induced intensity error is therefore
\begin{equation}
\delta I_P^\pm
=
I_0
\left[
\sqrt{2}r\sin\theta
\sin(\theta\pm2\mathcal{D})\cos\delta
+
r^2\sin^2\theta
\right].
\label{eq:stray_error_PeWVA}
\end{equation}

For SI, we define the normalized ideal output intensity as
\begin{equation}
S_{\mathrm{SI}}^\pm
=
\frac{I_{\mathrm{SI}}^\pm}{I_0}
=
\frac{1}{2}
\left[
1\pm\sin(2\theta)\sin(4\mathcal{D})
\right].
\end{equation}
Taking the ideal signal field as the phase reference, its normalized field amplitude can be written as $\sqrt{S_{\mathrm{SI}}^\pm}$. Under the same effective stray-field model, the detected intensity becomes
\begin{align}
\widetilde I_{\mathrm{SI}}^\pm
&=
I_0
\left|
\sqrt{S_{\mathrm{SI}}^\pm}
+
r e^{i\delta}
\right|^2
\nonumber\\
&=
I_{\mathrm{SI}}^\pm
+
I_0
\left[
2r\sqrt{S_{\mathrm{SI}}^\pm}\cos\delta
+
r^2
\right].
\label{eq:stray_SI}
\end{align}
Thus, the corresponding SI intensity error is
\begin{equation}
\delta I_{\mathrm{SI}}^\pm
=
I_0
\left[
2r\sqrt{S_{\mathrm{SI}}^\pm}\cos\delta
+
r^2
\right].
\label{eq:stray_error_SI}
\end{equation}

To compare the influence of the same optical perturbation on the estimation of $\theta$, we convert the intensity error into an equivalent optical-rotation error. If the perturbed intensity $\widetilde I$ is interpreted using the ideal calibration curve $I(\theta)$, then, for a sufficiently small perturbation,
\begin{equation}
\Delta\theta_s
\simeq
\frac{\widetilde I-I}
{\partial I/\partial\theta}
=
\frac{\delta I}
{\partial I/\partial\theta}.
\label{eq:stray_equivalent_error}
\end{equation}
The equivalent error is therefore governed by both the stray-light-induced intensity perturbation and the local slope of the transfer function. In PeWVA, postselection modifies the parasitic-field contribution while the steeper local response near the operating point increases the conversion from intensity to rotation, thereby reducing the resulting $\Delta\theta_s$.

\begin{figure}[!]
\includegraphics[width=\linewidth]{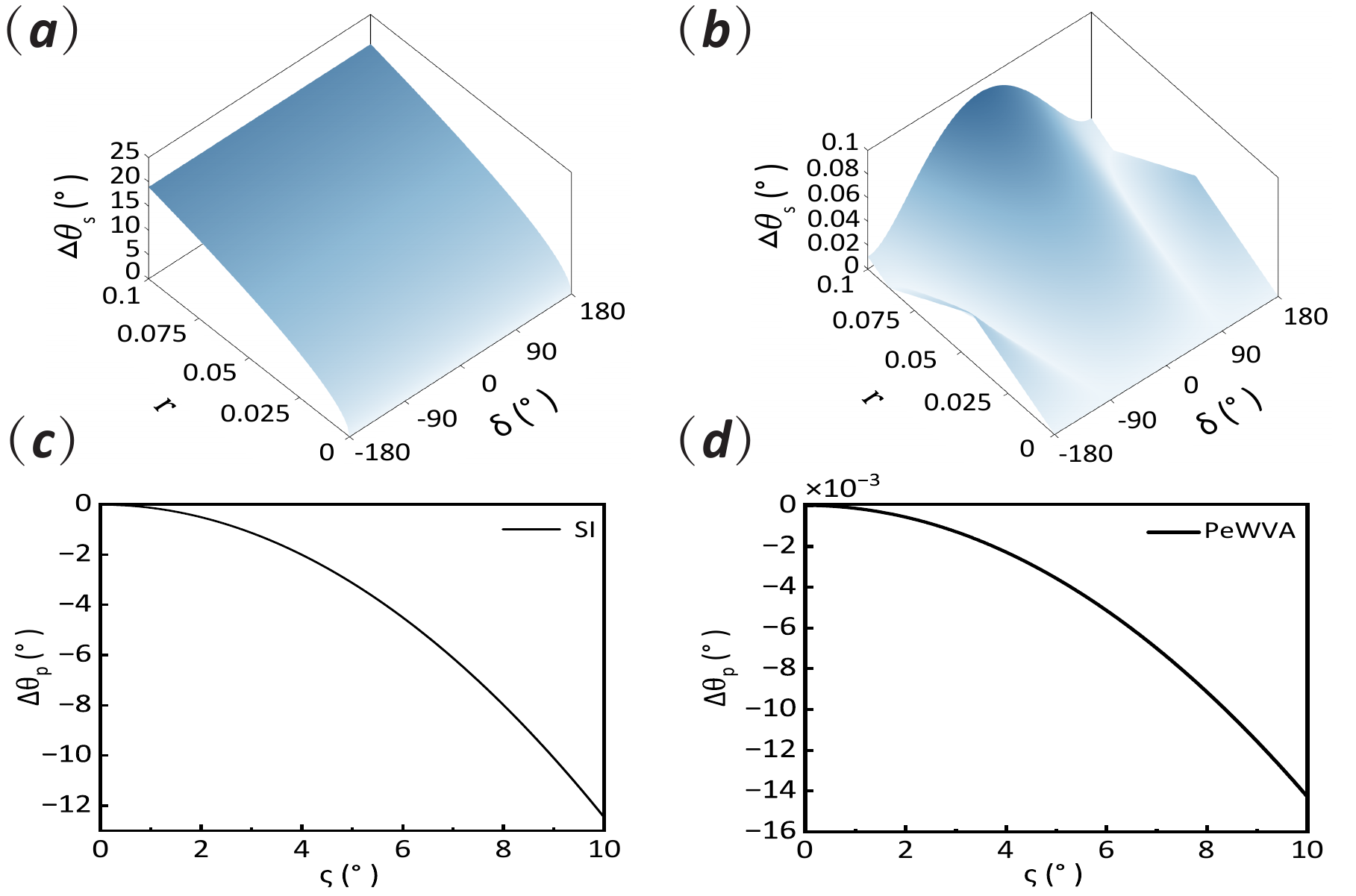}
\caption{Calculated equivalent OR errors for SI and PeWVA under representative interferometric perturbations. (a) and (b) Equivalent rotation error $\Delta\theta_s$ induced by coherent stray light as a function of the relative stray-field amplitude $r$ and phase $\delta$ for SI and PeWVA, respectively. (c) and (d) Equivalent OR error induced by relative interferometer-phase fluctuations for SI and PeWVA, respectively. The parameters are $\theta=45^\circ$ and $\mathcal D=22.5^\circ$ for SI, and $\theta=0.64^\circ$ and $\mathcal D=0.3^\circ$ for PeWVA. 
} 
\label{Fig.stray light}
\end{figure}

Figures~\ref{Fig.stray light}(a) and \ref{Fig.stray light}(b) show the calculated equivalent optical-rotation error as a function of the relative stray-field amplitude $r$ and phase $\delta$ for SI and PeWVA, respectively. For both
schemes, the error grows with increasing $r$ and exhibits a pronounced phase dependence due to coherent interference between the signal and stray fields. The PeWVA result, however, remains markedly smaller over the same range of
$r$ and $\delta$, indicating that the parasitic field is converted less efficiently into an apparent optical-rotation signal. 

\subsubsection{Interferometer phase jitter}
The relative phase between the two interferometer arms is another important source of technical error. Path-length drift, laser-frequency fluctuations, imperfect mode overlap, thermal deformation, and residual phase-controller errors can all perturb the interference condition
\cite{zhang2021investigation,badami2013displacement,
salvade2000limitations}. Such fluctuations modify the
interference term and can therefore appear as a spurious change in the inferred optical-rotation angle.
 
Let $\varsigma$ denote the relative phase between the two interferometer arms.
For SI, the output intensity at the port considered here is
\begin{equation}
I_{\mathrm{SI}}(\varsigma)
= \frac{I_0}{2} \left[1+\sin(4\mathcal{D})
\sin(2\theta) \cos\varsigma \right].
\label{eq:phase_SI}
\end{equation}
After postselection, the corresponding PeWVA intensity is
\begin{align}
I_P(\varsigma)
= \frac{I_0}{2}\Big[
&\cos^2(2\mathcal{D})\sin^2\theta
+ \sin^2(2\mathcal{D})\cos^2\theta
\nonumber\\
&+ \frac{1}{2}\sin(4\mathcal{D})
\sin(2\theta)\cos\varsigma \Big].
\label{eq:phase_PeWVA}
\end{align}
We take $\varsigma_0=\pi$ as the nominal dark-port phase and define the
phase-induced intensity variation as
$\delta I_{\varsigma}=I(\varsigma)-I(\varsigma_0)$. From
Eqs.~(\ref{eq:phase_SI}) and (\ref{eq:phase_PeWVA}), 
\[ \delta I_{\varsigma,\mathrm{SI}} = \frac{I_0}{2}
\sin(4\mathcal{D})\sin(2\theta)
(\cos\varsigma+1), \]
while \[
\delta I_{\varsigma,P} = \frac{I_0}{4}
\sin(4\mathcal{D})\sin(2\theta)
(\cos\varsigma+1).
\]
For a small phase excursion around the dark-port condition,
$\varsigma=\pi+\Delta\varsigma$, one has $\cos(\pi+\Delta\varsigma)+1\simeq\Delta\varsigma^2/2$. The phase-induced
intensity variation is therefore second order in small deviations from the nominal phase.

To compare the influence of this perturbation on the inferred optical rotation, we convert the intensity variation into an equivalent rotation error using the local response slope. Since SI and PeWVA operate at different values of
$\theta$ and $\mathcal{D}$, the equivalent rotation error provides an
appropriate basis for comparing their phase sensitivity at their respective
operating points. For a sufficiently small perturbation,
\begin{equation}
\Delta\theta_{\varsigma}(\varsigma)
\simeq
\frac{\delta I_{\varsigma}(\varsigma)}
{\left.\partial I/\partial\theta\right|_{\varsigma_0}}.
\label{eq:phase_equivalent_error}
\end{equation}
At $\varsigma_0=\pi$, the two local slopes are
\[
\left.
\frac{\partial I_{\mathrm{SI}}}{\partial\theta}
\right|_{\pi}
=
-I_0\sin(4\mathcal{D})\cos(2\theta),
\]
and
\[\left.
\frac{\partial I_P}{\partial\theta}
\right|_{\pi}
= \frac{I_0}{2}
\left[
\cos(4\mathcal{D})\sin(2\theta) -
\sin(4\mathcal{D})\cos(2\theta)\right].
\]
Together with the corresponding phase-induced intensity variations, these
local slopes determine $\Delta\theta_{\varsigma}$ for SI and PeWVA at their respective operating points.

The calculated results are shown in Figs.~\ref{Fig.stray light}(c) and
\ref{Fig.stray light}(d). As the interferometer phase deviates from the
dark-port condition, the equivalent rotation error increases for both
schemes. The substantially smaller error obtained with PeWVA originates
primarily from its much steeper local response to $\theta$ at the selected
operating point. According to Eq.~(\ref{eq:phase_equivalent_error}), a larger
$\partial I/\partial\theta$ converts a given phase-induced intensity
perturbation into a smaller equivalent optical-rotation error. The
postselection also modifies the phase-dependent interference term, further
affecting $\delta I_{\varsigma}$. 

\subsection{Atomic Faraday rotation as preselection encoding}
 
The preceding framework treats $\theta$ as a generic parameter-dependent polarization angle and therefore does not rely on a particular sensing mechanism. In the atomic implementation used here, this angle is generated by the dispersive interaction between the probe polarization and the collective
atomic spin. For a far-detuned linearly polarized probe, the vector component of the atomic polarizability gives rise to the standard dispersive Faraday interaction
\cite{deutsch2010quantum,vasilyev2012quantum}
\begin{equation}
H_F=\gamma S_3\left(\mathbf{j}\cdot\hat{\mathbf n}\right),
\label{eq:Faraday_Hamiltonian}
\end{equation}
where $S_3$ is the optical Stokes operator, $\mathbf{j}$ is the collective
ground-state angular momentum, and $\hat{\mathbf n}$ is the probe propagation
direction. The two circular components of the probe acquire opposite
spin-dependent phase shifts during propagation through the vapor, producing a
rotation of the linear polarization. The probe therefore carries an optical
record of the collective spin projection
$J_n\equiv\langle\mathbf{j}\cdot\hat{\mathbf n}\rangle$, which we write as
$\theta_{\rm at}=\kappa J_n$, where $\kappa$ contains the atomic density, cell length, transition strengths, and probe detuning.

With an experimentally introduced polarization bias $\theta_0$, the atomic ensemble prepares the state entering the PeWVA interferometer as
\begin{equation}
\ket{\chi(J_n)}
= \sin\theta\,\ket{H}+\cos\theta\,\ket{V},
\qquad
\theta=\theta_0+\kappa J_n.
\label{eq:atomic_state_mapping}
\end{equation}
This relation establishes the connection between the atomic sensor and the general PeWVA framework: the collective spin projection is encoded directly in the preselected polarization state, before the fixed weak interaction
and postselection are applied. Consequently, the weak value becomes an
explicit function of the collective spin projection,
\begin{equation}
A_w(J_n)
=-i\cot\!\left(\theta_0+\kappa J_n\right).
\label{eq:atomic_weak_value}
\end{equation}
Around an operating point $J_{n,0}$, with
$J_n=J_{n,0}+\delta J_n$ and
$\bar\theta=\theta_0+\kappa J_{n,0}$, the corresponding weak-value variation is
\[
\delta A_w\simeq i\kappa\,\csc^2\!\bar\theta\;\delta J_n.
\]
Thus, a change in the collective spin projection changes the overlap between
the preselected and postselected states and is mapped onto a variation of the weak-value readout. 

For the atomic magnetometer used in this work, the magnetic field drives the collective-spin dynamics and changes the spin projection sampled by the probe. The resulting variation of $J_n$ is first converted by the atomic ensemble into Faraday rotation and is then read out by the PeWVA interferometer. Near a
chosen magnetic operating point, this combined transduction may be written to
first order as
\[
\delta\mathcal V_P
\simeq
\left.
\frac{\partial\mathcal V_P}{\partial\theta}
\right|_{\bar\theta_P}
\kappa
\left.
\frac{\partial J_n}{\partial B}
\right|_{\bar B}
\delta B.
\]
A small variation of the magnetic field first perturbs the collective-spin projection, with the response characterized by $\partial J_n/\partial B$. The resulting change in the spin-dependent Faraday rotation is then mapped onto the optical readout through the local response $\partial\mathcal V_P/\partial\theta$. PeWVA acts at this optical readout stage, increasing the response to the encoded spin signal.

This expression separates the atomic magnetic response,
$\partial J_n/\partial B$, from the optical readout response,
$\partial\mathcal V_P/\partial\theta$. PeWVA leaves the former unchanged but
enhances the latter near the selected operating point. Accordingly, after
calibration of $\kappa$ and the atomic response, the measured postselected
signal can be used to infer the collective spin projection and, in the
magnetometer, the magnetic field that generates it. This separation also makes
clear that the atomic ensemble serves as the physical parameter encoder,
whereas the downstream interferometer provides the fixed weak coupling and
postselected readout. A microscopic derivation of the alkali polarizability is
not required for the PeWVA framework itself and can be retained in the
Supplemental Material.

\section{EXPERIMENTAL DEMONSTRATION WITH AN ATOMIC MAGNETOMETER}

We experimentally demonstrate PeWVA using an $^{87}$Rb $M_x$ atomic magnetometer, in which the magnetic-field-dependent collective-spin response is encoded onto the probe polarization through Faraday rotation. The experiment provides a direct realization of the preselection-encoding mechanism introduced in Sec.~II and allows us to characterize the readout performance under realistic operating conditions, including detector saturation and optical imperfections. In addition, controlled interferometric perturbations are investigated to further validate the robustness and applicability of the PeWVA scheme.

\subsection{Experimental implementation}

\begin{figure}[t]
\includegraphics[width=\linewidth]{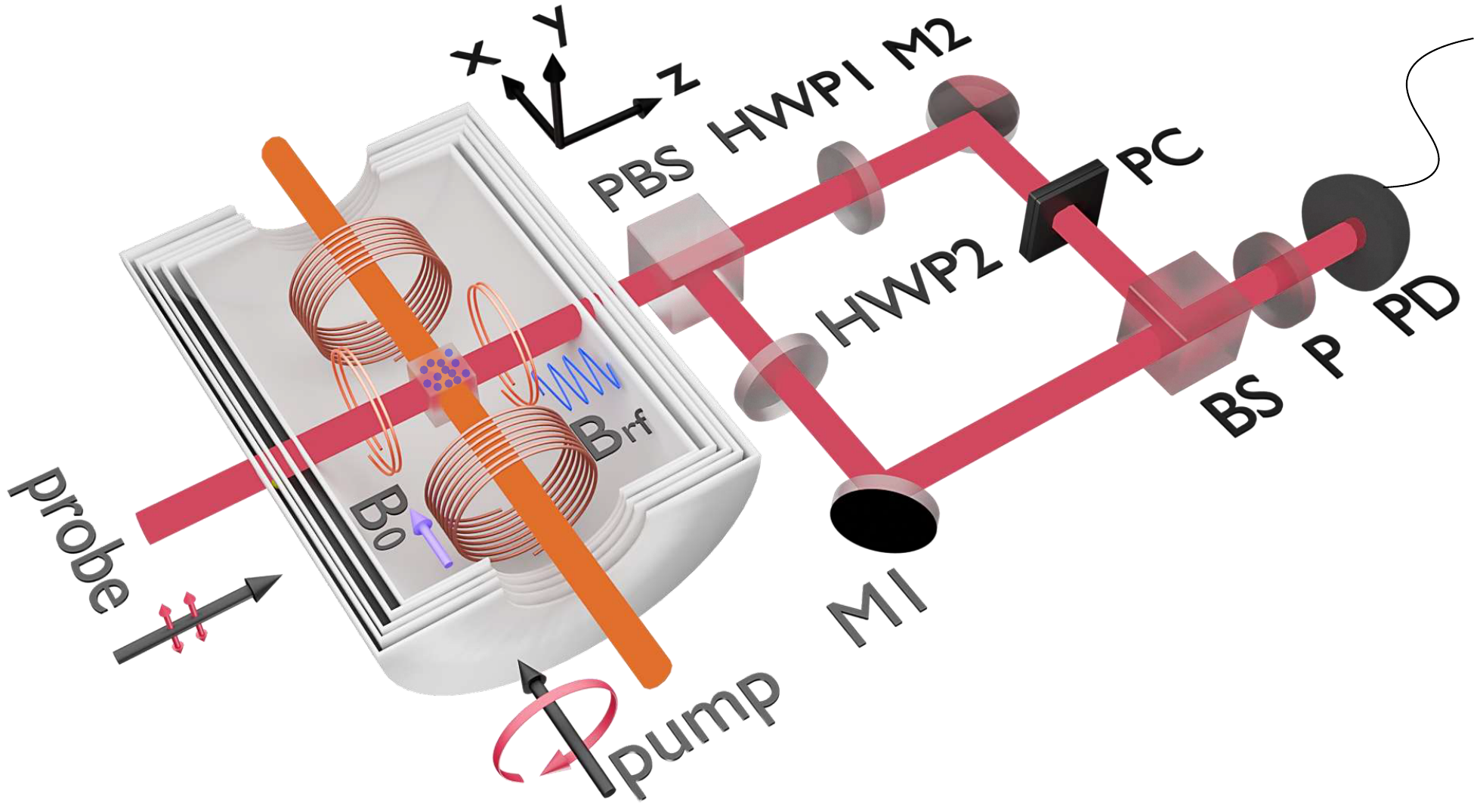}
\caption{Experimental configuration of the $^{87}$Rb $M_x$ atomic magnetometer and downstream PeWVA readout. The vapor cell is located inside a five-layer magnetic shield and is addressed by orthogonal pump and probe beams. The static field $B_0$ is applied along the pump direction and the rf field $B_{\mathrm{rf}}$ along the probe direction. After the cell, the probe enters the polarization interferometer, where HWP1 and HWP2 introduce opposite weak rotations $\pm\mathcal D$, the phase controller sets the relative arm phase, and the final polarizer performs the postselection. Removing the postselection polarizer realizes the SI reference. HWP,  half-wave plate; P, polarizer; PBS, polarizing beam splitter; M, mirror;  PC, phase controller; BS, beam splitter; APD, avalanche photodiode. } 
\label{layout}
\end{figure}

The experimental arrangement is shown in Fig.~\ref{layout}. The sensing element is a $3\times3\times3~\mathrm{mm^3}$ cubic vapor cell containing enriched $^{87}$Rb and 500 Torr of $\mathrm{N_2}$. The cell is mounted inside a five-layer magnetic shield that suppresses geomagnetic and ambient magnetic fields.  Three mutually orthogonal Helmholtz-coil pairs provide the static and oscillating fields required for operation of the magnetometer. The static bias field $B_0$ is applied along the $x$ direction, parallel to the pump beam, while the radio-frequency (rf) field $B_{\mathrm{rf}}$ is applied along the $z$ direction, parallel to the probe beam. In the experiment, the bias field is set to $B_0=1000~\mathrm{nT}$, corresponding to a Larmor frequency of approximately 7 kHz, which is used as the resonant frequency of the rf driving field. The vapor cell is maintained at around 373 K using a twisted-pair resistive heater wrapped around a cubic boron-nitride holder. The heater is driven at 200 kHz to move the heating current well above the magnetic-signal band. A circularly polarized pump beam produced by a distributed-feedback laser is tuned to the $^{87}$Rb D$_1$ transition and propagates along $x$, with an optical power of $180~\text{\textmu W}$ at the vapor cell. A linearly polarized probe beam from an external-cavity diode laser propagates along the $z$ axis, has a diameter of approximately 3 mm, and is blue-detuned by 162.515 GHz from the $^{87}$Rb D$_2$ transition. The probe power is varied in the detector-saturation measurements described below. Under these conditions, the driven transverse atomic-spin component produces a time-dependent Faraday rotation
of the probe at the magnetic resonance.

Before the probe enters the vapor cell, a half-wave plate (HWP) introduces an adjustable polarization bias $\theta_0$. The polarization angle after the atomic interaction is therefore given by the sum of this static bias and the spin-dependent Faraday rotation, such that $\theta=\theta_0+\theta_{\mathrm{at}}$. The bias angle determines the operating point of the preselected state on the PeWVA transfer function. As illustrated in Fig.~\ref{simulation results}, for a fixed $\mathcal D$, an operating point closer to the nearly orthogonal pre- and postselection condition provides a steeper local response to small variations of $\theta$, while simultaneously reducing the usable linear range. We therefore choose $\theta_0$ on the steep, approximately linear part of the PeWVA response, with its specific value determined by the required compromise between local amplification and dynamic range. For the detector-performance measurements described below, we use $\mathcal D=1^\circ$ and $\theta_0=0.64^\circ$, which provide a practical balance between response enhancement and linear measurement range.

After leaving the cell, the probe enters the polarization interferometer. A polarizing beam splitter separates the two linear-polarization components, and HWPs in the two arms introduce the opposite rotations $\pm\mathcal D$. A phase controller sets the relative interferometer phase close to an odd multiple of $\pi$ so that the selected output operates near the dark-port condition. The two paths are recombined at a 50/50 beam splitter, and a final polarizer performs the postselection in PeWVA. Removing this polarizer realizes the SI reference without changing the upstream atomic sensing stage. The optical output is measured using either a PD (Thorlabs PDA100A2) operated at the 40-dB gain setting or an APD (Thorlabs APD430A2) operated near its maximum avalanche gain (\(M\approx100\)). The corresponding nominal linear detection ranges are $120$ $\text{\textmu W}$ and $8$ $\text{\textmu W}$, respectively. The oscillating detector signal is demodulated with a lock-in amplifier at the Larmor frequency to obtain the response used for magnetic readout, denoted by $Q$. 

\begin{figure}[t]
\centering
\includegraphics[width=0.8\linewidth]{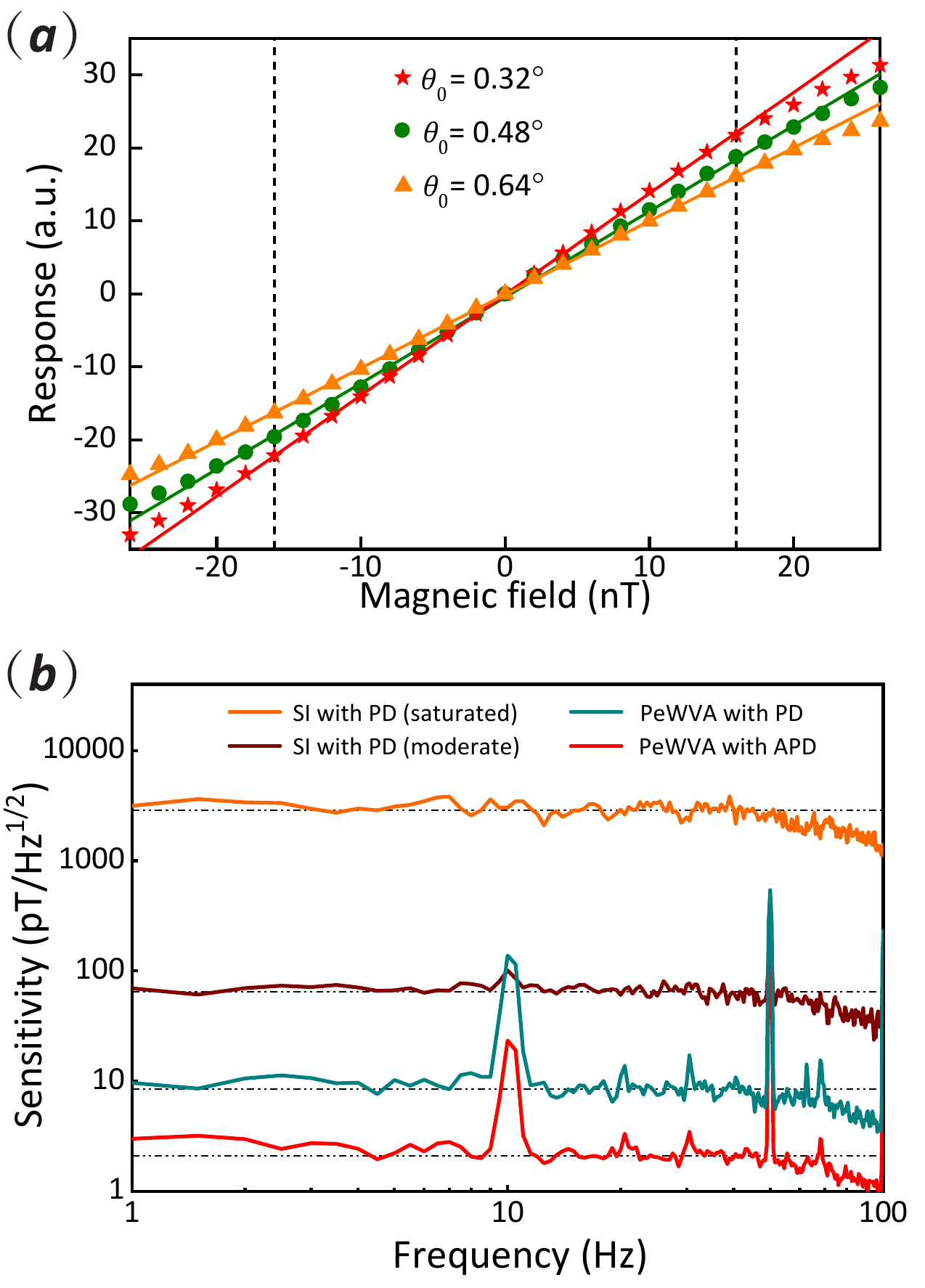}
\caption{Magnetic-field calibration and detector-constrained noise performance. (a) Demodulated response $Q$ as the static magnetic field is scanned over $\pm26~\mathrm{nT}$ around $B_0=1000~\mathrm{nT}$ for $\theta_0=0.32^\circ$, $0.48^\circ$, and $0.64^\circ$, with $\mathcal D=1^\circ$. The dashed vertical lines indicate the approximate linear range for $\theta_0=0.32^\circ$. (b) Magnetic-field noise spectra for saturated SI with a PD, unsaturated SI with reduced probe power, PeWVA with a PD, and PeWVA with an APD. The dash-dotted lines indicate the representative noise floors discussed in the text.}
\label{Fig.experimental_results}
\end{figure}

Prior to each measurement, the interferometer phase is adjusted to the dark-port operating condition, and the magnetometer response is calibrated by scanning the static field around $B_0=1000~\mathrm{nT}$. Figure~\ref{Fig.experimental_results}(a) shows representative calibration curves obtained by scanning the field over $\pm26~\mathrm{nT}$. In these measurements, the weak-coupling angle is fixed at $\mathcal D=1^\circ$, while the polarization bias is varied among $\theta_0=0.32^\circ$, $0.48^\circ$, and $0.64^\circ$. Reducing $\theta_0$ increases the local calibration slope $dQ/dB$, but simultaneously narrows the field range over which the response remains approximately linear. For $\theta_0=0.32^\circ$, the response remains approximately linear over a magnetic-field range of $\pm15~\mathrm{nT}$ around the operating point, with a fitted local slope of 1.384~$\mathrm{mV/nT}$. Beyond this range, the deviation from linearity is dominated by the nonlinear PeWVA transfer function, with a smaller contribution from the atomic-magnetometer dispersion. These measurements experimentally confirm the trade-off between gain and dynamic range predicted by the PeWVA transfer function. Operation closer to the nearly orthogonal pre- and postselection condition produces a larger response to the spin-dependent Faraday rotation, but over a narrower usable range.

\subsection{Detector-constrained magnetic-field readout}

We next compare the magnetic-field noise obtained with SI and PeWVA under different detector conditions. For the PeWVA measurements in this comparison, we use $\mathcal D=1^\circ$ and $\theta_0=0.64^\circ$. The SI reference is operated at $\mathcal D=22.5^\circ$ and $\theta_0=45^\circ$, consistent with the operating point adopted in the theoretical comparison. Magnetic-field records were acquired 30 times at intervals of 2 s. Each detector-voltage record was converted to magnetic field using the locally measured $dQ/dB$, and the noise spectra were obtained by Fourier transformation and averaging. 

Figure~\ref{Fig.experimental_results}(b) compares four representative operating conditions. With SI and a PD, an incident probe power of 4 mW drives the detector outside its linear response range and produces a magnetic-field noise floor of approximately $2870~\mathrm{pT/\sqrt{Hz}}$. Reducing the probe power to 0.37 mW restores unsaturated SI operation, but the lower optical signal level results in a noise floor of approximately $65~\mathrm{pT/\sqrt{Hz}}$. In PeWVA, the same type of PD can be operated with an incident probe power of 4 mW, while postselection strongly attenuates the power delivered to the detector, yielding a noise floor of approximately $8.5~\mathrm{pT/\sqrt{Hz}}$. Replacing the PD with an APD further reduces the measured floor to approximately $2.1~\mathrm{pT/\sqrt{Hz}}$ at an incident probe power of 1.4 mW. 

These measurements show the practical consequence of decoupling the probe power used at the atomic sensing stage from the optical power received by the detector. In SI, increasing the incident power eventually produces detector nonlinearity, whereas reducing the power to recover linear operation degrades the optical SNR. In PeWVA, the atomic ensemble can be interrogated with a substantially larger probe flux while postselection keeps the detected flux within the useful range of the receiver. The same mechanism also permits the use of the APD, whose higher gain is advantageous at low detected power but whose usable optical-power range is more restrictive. The measured improvement should therefore be interpreted as a detector-constrained readout advantage rather than as an increase in the information carried by each incident photon, consistent with the Fisher-information analysis of Sec.~II. A narrow spectral feature near 10 Hz is observed in the measured spectra and is traced to the phase-controller electronics rather than to the atomic sensor itself.

\subsection{Technical perturbations}

\begin{figure}[t]
\includegraphics[width=0.8\linewidth]{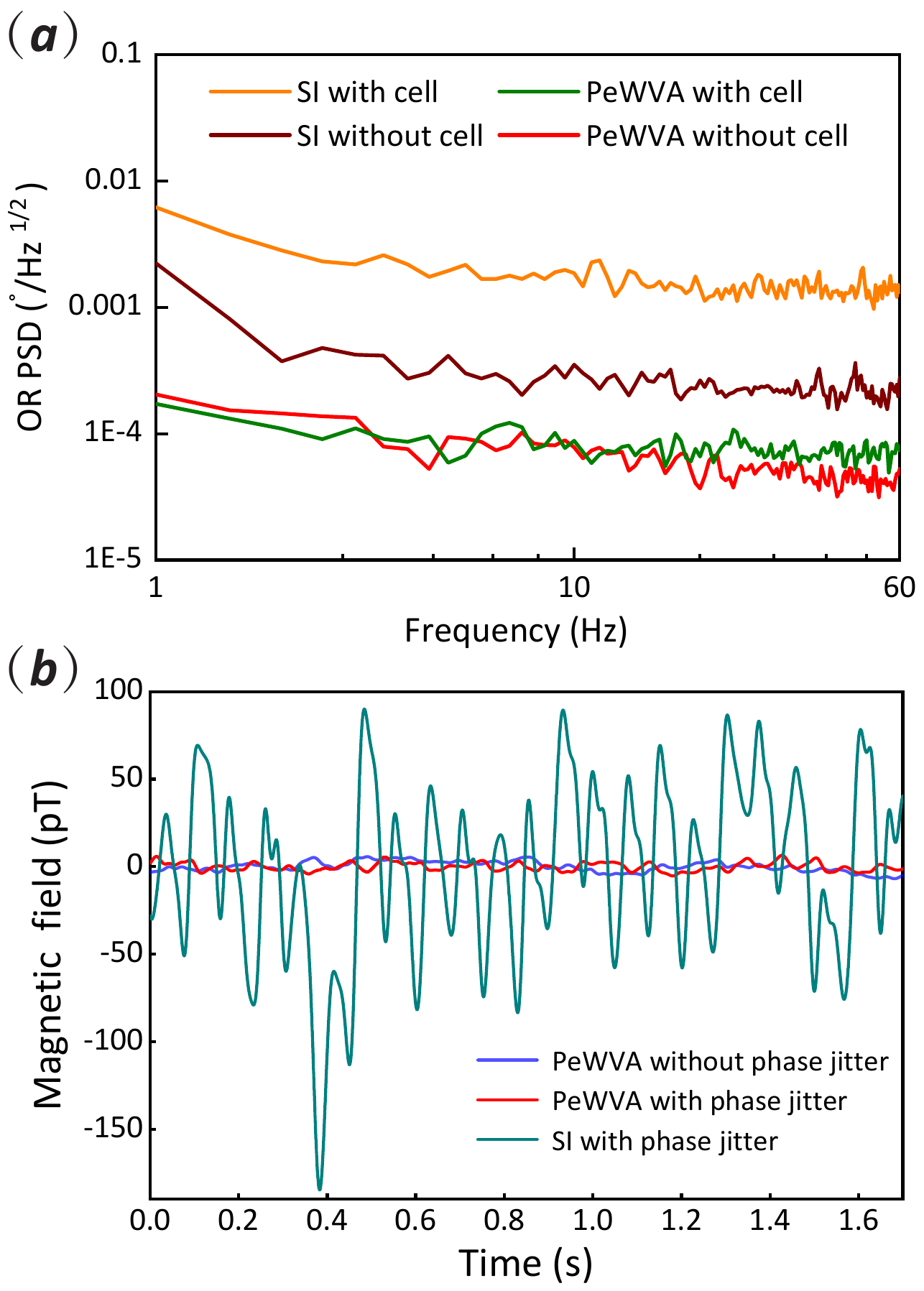}
\caption{Experimental response to representative interferometric perturbations. (a) OR noise spectra measured with and without the vapor cell for SI and PeWVA. Insertion of the cell reduces the interferometer visibility from 0.96 to 0.63, while the resulting degradation is substantially smaller for PeWVA than for SI. (b) Magnetic-field-equivalent response under a controlled 15-Hz sawtooth modulation of the relative interferometer phase with an amplitude of 0.024 probe wavelengths. The PeWVA trace without imposed phase modulation is included as a reference.} 
\label{Figphasejitter}
\end{figure}

The separation of the atomic sensing stage from the downstream interferometer also enables a direct experimental test of two technical perturbations considered in Sec.~II. We first examine coherent stray light originating from the vapor cell. Multiple reflections from the cell windows are significant in the present apparatus: the interferometer visibility decreases from 0.96 when the cell is removed to 0.63 when the cell is inserted. Figure~\ref{Figphasejitter}(a) compares the equivalent optical-rotation noise measured with and without the cell for SI and PeWVA under their respective optimized operating conditions. In SI, insertion of the vapor cell leads to a pronounced increase in the measured noise spectrum, whereas the corresponding degradation is substantially suppressed in PeWVA. Quantitatively, the noise floor increases by factors of 5.13 and 6.23 for SI within the 1--10 Hz and 10--20 Hz bands, respectively, while the corresponding factors for PeWVA are only 0.98 and 1.19.
This behavior is consistent with the model in Sec.~II. Although the parasitic cell field is not removed, its conversion into an equivalent rotation error is reduced at the PeWVA operating point because the optical response to the encoded Faraday rotation is much steeper.

We next impose a controlled phase perturbation using the phase controller in one interferometer arm. The relative phase is modulated with a 15-Hz sawtooth waveform with an amplitude of 0.024 probe wavelengths while the background field is maintained at $B_0=1000~\mathrm{nT}$. Figure~\ref{Figphasejitter}(b) compares the magnetic-field-equivalent response induced by the imposed phase modulation for SI and PeWVA. The imposed phase modulation produces a much larger spurious response in SI than in PeWVA at their respective operating points. The reduced error primarily arises from the larger local optical response to the encoded rotation in PeWVA. For a given phase-induced intensity perturbation, the steeper calibration slope corresponds to a smaller inferred rotation error. Postselection also modifies the phase-dependent interference term, as discussed in Sec.~II.

\section{Conclusion and Outlook}

We have introduced preselection-encoded weak-value amplification as a polarization-interferometric readout architecture for optical-rotation measurements. In this scheme, the physical sensing process first encodes the quantity of interest into the probe polarization, and a downstream interferometer subsequently applies a fixed weak coupling and postselection. Moving the parameter encoding into the preselected state separates the sensing interaction from the phase-sensitive weak-value readout, allowing the local optical response to be tuned through the downstream interferometer without modifying the sensing process itself.

The theoretical analysis establishes three key features of PeWVA. First, reducing the weak-coupling angle enhances the local response near the operating point at the cost of a reduced usable dynamic range. Second, the Fisher-information analysis shows that postselection does not create additional information at fixed incident photon number, but redistributes the available information into a lower-power detected channel. The practical advantage
therefore emerges when the detector becomes the limiting resource.  Third, we
derive the response of PeWVA to two representative interferometric perturbations, coherent stray light and relative phase fluctuations. In both cases, the perturbation-induced intensity variation is converted into an equivalent
optical-rotation error. The analysis shows that, at the operating points considered
here, the steeper PeWVA response and the modified interference term reduce the corresponding equivalent errors
relative to standard interferometric readout.

These predictions are verified experimentally using an $^{87}$Rb $M_x$ atomic magnetometer. The measured calibration curves reproduce the predicted trade-off between enhanced local response and reduced dynamic range. Under detector-constrained operation, PeWVA achieves a magnetic-field noise floor of approximately $8.5~\mathrm{pT}/\sqrt{\mathrm{Hz}}$ with a conventional photodiode, substantially below the $65~\mathrm{pT}/\sqrt{\mathrm{Hz}}$ obtained with unsaturated standard-interferometric readout at reduced probe power. Replacing the photodiode with an avalanche photodiode further lowers the measured noise floor to approximately $2.1~\mathrm{pT}/\sqrt{\mathrm{Hz}}$. The response to these two representative interferometric perturbations is also examined experimentally. When the vapor cell is inserted, coherent reflections from the cell windows produce a much larger degradation in the SI readout than in PeWVA. Likewise, under a controlled modulation of the relative interferometer phase, the resulting spurious magnetic-field response is substantially smaller for PeWVA. These results establish PeWVA as a flexible optical readout strategy that combines tunable local amplification, reduced detected optical
power, and reduced conversion of specific interferometric perturbations into measurement error, without modifying
the underlying atomic response.

Because PeWVA only requires the sensing stage to prepare a parameter-dependent polarization state before the downstream interferometer, the same architecture can in principle be extended beyond atomic magnetometry to a broad class of optical-rotation and polarization-encoded measurements. Possible examples include optical-rotation spectroscopy of chiral molecules, Faraday-rotation diagnostics of magnetized plasmas, Faraday-rotation spectroscopy for trace-gas sensing, and astronomical measurements in which Faraday rotation is used to probe magnetic fields in the interstellar and intergalactic media \cite{delage2025chiral,kondru1998atomic,lewicki2009ultrasensitive,kaczmarek2017detection,michilli2018extreme,gaensler2005magnetic}. More broadly, PeWVA provides a platform for exploring how weak-value readout can be combined with additional quantum resources. Future efforts may investigate photon recycling to recover information carried by rejected photons and improve photon utilization and measurement precision \cite{lyons2015power,wang2016experimental}, as well as squeezed or entangled optical probes to extend the measurement precision beyond the coherent-state shot-noise limit \cite{novikov2025hybrid,ligo2011gravitational,wu2023quantum}. Such developments may open a route toward integrating detector-constrained weak-value readout with genuinely quantum-enhanced optical metrology.

\section*{Acknowledgments}
This work was supported by Natural Science Foundation of Zhejiang Province (Grant No. LY24A050005) and the National Natural Science Foundation of China (Grant Nos. 12574532 and 92476204).

 \bibliographystyle{naturemag}

\end{document}